%% file: HiggsTevatron.tex
\documentclass[12pt]{article}
\usepackage{epsfig}
\usepackage{graphics}
\input econfmacros.tex
\def\Title#1{\begin{center} {\Large {\bf #1} } \end{center}}

\newcommand {\ra}         {\rightarrow}
\newcommand{\GeV}{\mbox{$\mathrm{GeV}$}}

\newcommand{\LepII}{\mbox{LEP 2}}

\newcommand{\MET}{$/\!\!\!\!E_{T}$}

\newcommand {\ttbar}      {t \bar t}
\newcommand {\bbbar}      {b \bar b}
\newcommand {\qqbar}      {q \bar q}
\newcommand {\ppbar}      {p \bar p}

\def\PR#1#2#3{{\rm Phys.~Rept.} {\bf#1} (#2) #3}
\def\PLB#1#2#3{{\rm Phys.~Lett.} {\bf{B#1}} (#2) #3}
\def\NPB#1#2#3{{\rm Nucl.~Phys.} {\bf{B#1}} (#2) #3}
\def\PRL#1#2#3{{\rm Phys.~Rev.~Lett.} {\bf{#1}} (#2) #3}

\begin{document}
\Title{Higgs Searches at the Tevatron
\footnote{Presented at the Symposium ``Physics in Collision'', Perugia, June 25-28, 2008.}}
\medskip

\begin{center} 
\Large {Anyes Taffard }\\
\small {University of California Irvine, CA 92697, U.S.A.}
\end{center}
\begin{center} 
{\it on behalf of the CDF and D\O~Collaborations }
\end{center}

\abstract 
We review the status of the searches for the Higgs boson in the context of the 
Standard Model and the Minimal SuperSymmetric Standard Model by the
CDF and D\O~experiments at the Fermilab Tevatron proton-antiproton collider, 
using up to $2.4~\rm{fb}^{-1}$ of Run II data. Since no evidence of signal above
the expected background is observed in any of the various final states examined, 
limits at 95\% confidence level are presented.

\section{Introduction}

In the Standard Model (SM), the Higgs mechanism breaks the electroweak symmetry by 
introducing a scalar field to generate particle masses. It predicts the existence of 
a neutral spin 0 boson, the Higgs boson, but not its mass. Direct searches at \LepII~have 
excluded a SM Higgs boson with mass below $114.4~\GeV$ at 95\% confidence level (C.L.). 
Indirect measurements from SLD, LEP and the Tevatron, favor a light Higgs boson with 
mass of $87^{+36}_{-27}~\GeV$ at 68\% C.L. and constrain its mass to be below $190~\GeV$ 
at 95\% C.L. when including the \LepII~exclusion~\cite{ref:1}. 

At the Tevatron, SM Higgs production is dominated by gluon fusion, with smaller contributions
from W or Z bosons associated productions. Cross sections range from 0.1 to 1 pb. 
Below $135~\GeV$ (low-mass), the SM Higgs boson decays predominantly to $\bbbar$. In order to 
avoid the huge SM QCD multijets background, searches use the associated productions: $WH$ and $ZH$. 
Above $135~\GeV$ (high-mass), SM Higgs decays predominantly to $WW^*$, thus making it possible to 
use the gluon fusion production. 

Many models beyond the SM, including Supersymmetry, predict larger Higgs production cross
sections, some of which within reach in the present data sets~\cite{ref:2}. The Minimal Supersymmetric 
extension of the SM (MSSM) introduces two Higgs doublets separately coupling to up-type and down-type fermions.
Out of the eight degree of freedom, three result in the longitudinal components of the $W^{\pm}$ and Z bosons 
and the remaining in five physical Higgs bosons. Two of them are CP-even scalars ($h$, $H$), 
where $h$ is the lightest and SM-like. The other three consist of a charged Higgs pair ($H^{\pm}$) and 
a CP-odd scalar (A) the mass of which ($m_A$) is one of the two free parameters of the model at tree-level. 
In the MSSM, the Higgs production cross section is proportional to square of the second free parameter 
of the model, $\tan\beta$, the ratio of the two vacuum expectation values of the Higgs doublets. Thus,
large values of $\tan\beta$ result in significantly increased production cross section compared to the SM. 
Additionally, in the large  $\tan\beta$ limit, the heaviest CP-even Higgs boson, H, and the CP-odd Higgs 
scalar, A, are predicted to be almost degenerate in mass, leading to a further cross section enhancement.
The main production mechanism for neutral Higgs bosons ($\phi=h,H,A$) are $gg,\bbbar \ra \phi$ and 
$gg, \qqbar \ra \phi + \bbbar$, where the branching ratio of $\phi \ra \bbbar$ is around 90\% and 
$\phi \ra \tau^+ \tau^-$ is around 10\%. Due to the lower background in the $\tau$ channel, the overall
experimental sensitivity is similar in both channels. Other extension to the SM, such as Top-color~\cite{ref:3}
or Fermiophobic Higgs model~\cite{ref:4}, predict an enhanced branching fraction of neutral Higgs bosons to two photons.
Extensions of the Higgs sector involving higher isospin multiplets predict the existence of double-charged Higgs bosons, 
which can be relatively light and hence accessible at the Tevatron.

This document describes the searches for a SM and beyond SM Higgs boson by the CDF and D\O~collaborations using up
to $2.4~\rm{fb}^{-1}$ of Run II data. The majority of those results are preliminary, and more information
can be found on the public pages of CDF~\cite{ref:5} and D\O~\cite{ref:6}.

\section{Standard Model Higgs searches}

\subsection{Search for $ZH \ra l^+l ^-b \bar b$}

\begin{figure}[htb]
\begin{center}
\includegraphics[width=.3\textheight]{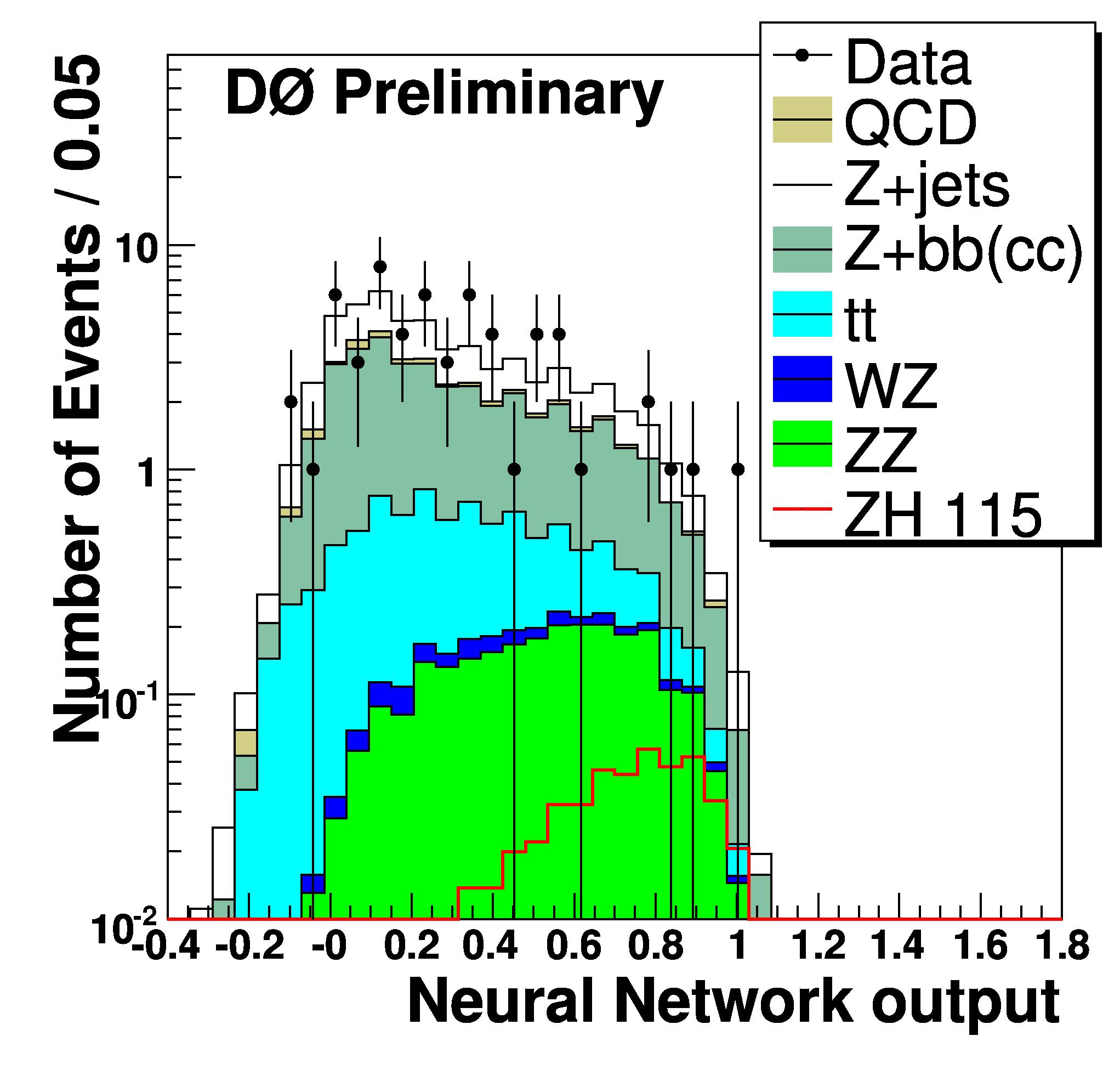}  
\includegraphics[width=.33\textheight]{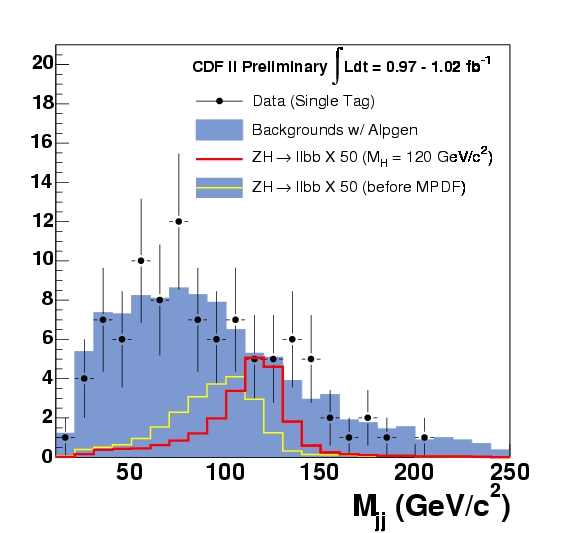}  
\caption{D\O~NN output for the double tag sample (left). 
CDF di-jet mass distribution after using the NN correction function (right). 
The 2 lines histograms show the effect of the correction on a $ZH$ signal for $M_H=120~\GeV$.}
\label{fig:1}
\end{center}
\end{figure}

For the low mass region (below $135~\GeV$) where the Higgs decays predominantly to $\bbbar$,  
the cleanest channel is the associated production with a Z boson, where the Z decays leptonically to 
$e^+e^-$ or $\mu^+\mu^-$. Although the cross section times branching ratio is lower that the associated 
production with a W boson, this channel offers several tight constraints since $M_{l^+l^-}=M_Z$ and the lack of direct
missing transverse energy (\MET) can be used to improve the jet energy resolution. Candidate events are selected 
by requiring two high $P_T$ (typically above $15~\GeV$) electrons or muons of opposite charge with invariant mass
matching that of a Z boson. D\O~requires two jets with $E_T>15~\GeV$, while CDF requires additionally 
that the highest energetic jet passes $E_T>25~\GeV$. After this pre-selection, the sample is dominated by Z+jets
and therefore $b$-jets identification is crucial to reduce this background. D\O~utilises an artificial 
neural network (NN) tagger  based on lifetime information, which performs with efficiencies 
ranging 50-70\% for a mis-identification (also referred to as mistag) rate of 0.3-4.5\%. 
CDF uses a secondary vertex reconstruction algorithm with efficiencies ranging 40-50\% for a mistag 
rate of 0.3-0.5\%. To further discriminate signal and background events, D\O~uses a NN separately 
trained for 1-tag and 2-tag events using ten kinematic variables (Figure~\ref{fig:1}). 
Since no excess of signal is observed, the output 
of the NN is fitted to extract a 95\% C.L. limit corresponding to 17.8 (20.4 expected) above the theoretical 
SM expectation for $M_H=115~\GeV$. 
CDF uses a NN to improve the dijet mass distribution, which essentially is a correction function that
re-assigns the~\MET~to the jets (Figure~\ref{fig:1}). Then, a 2-dimensional NN, trained to separate ZH from 
top background and ZH from Z+jets background, is used separately for single and double tags events.
The NN output is fitted to extract a 95\% C.L. limit corresponding to 16 (16 expected)  above the theoretical 
SM expectation~\cite{ref:7}. 

\subsection{Search for $ZH \ra \nu \bar \nu b \bar b$}

\begin{figure}[htb]
\begin{center}
\includegraphics[width=.33\textheight]{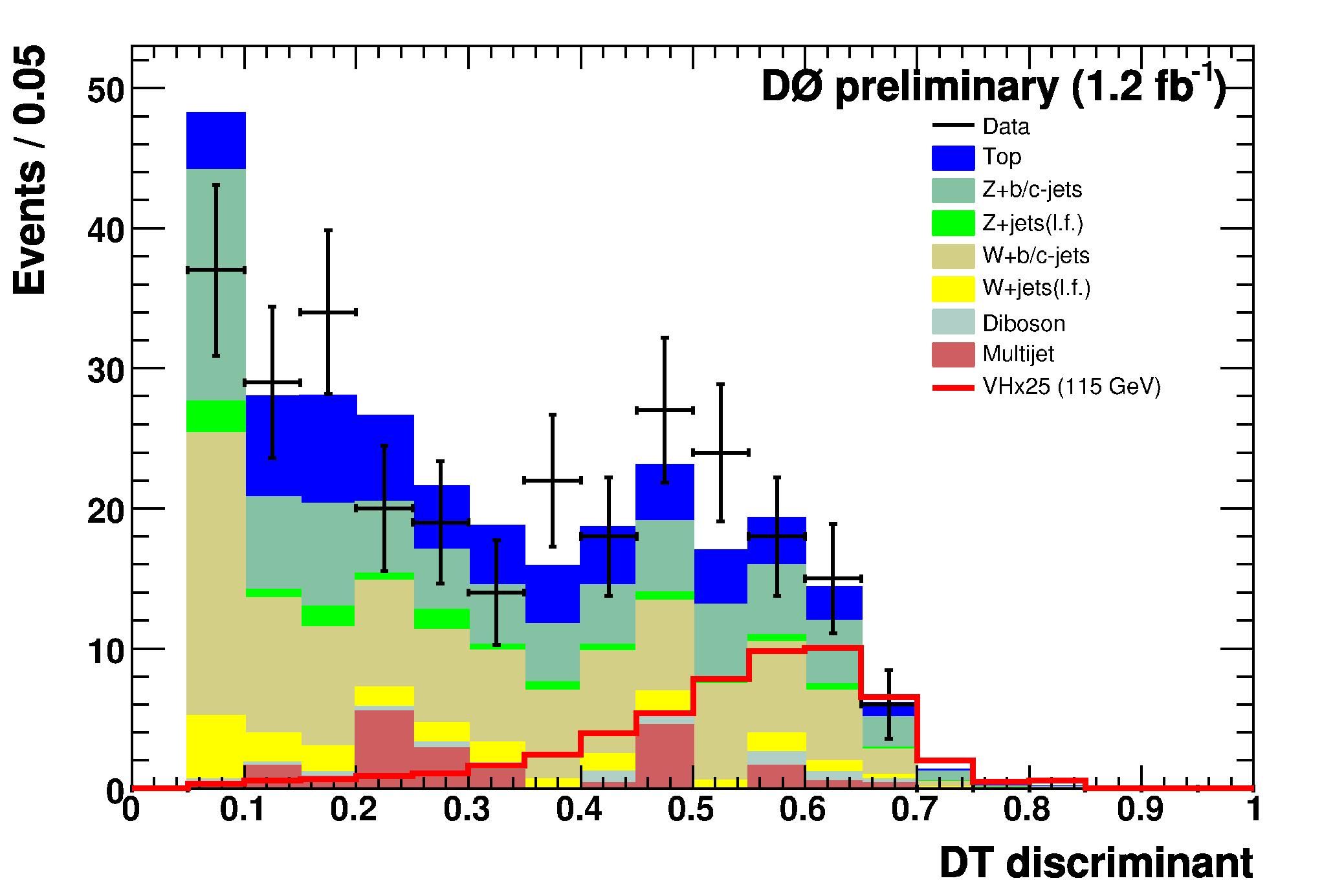}  
\includegraphics[width=.33\textheight]{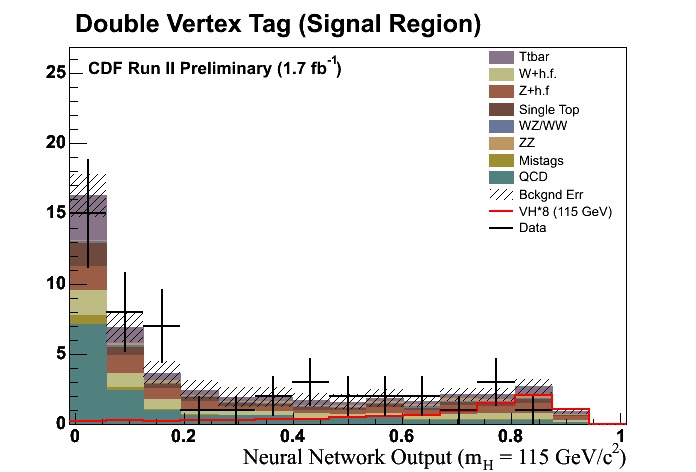}  
\caption{D\O~DT output (left) and CDF NN output (right) for the double tag sample.}
\label{fig:2}
\end{center}
\end{figure}

Although this channel has a larger branching ratio to neutrinos, it is very challenging to   
trigger on and background wise. Events are triggered on jets plus~\MET~and tight cuts are applied 
to reject background. D\O~requires two or three jets with $E_T>20~\GeV$, where the two leading ones are not back-to-back and 
with at least $50~\GeV$ of~\MET~in the event. CDF asks for two jets only, where the $E_T$ of the leading jet is above $45~\GeV$, 
$25~\GeV$ for the second jet and at least $50~\GeV$ of~\MET~not aligned with the jets. 
Both experiments require that there be no identifies electrons or muons so that the data sample 
is orthogonal to the $WH\ra l \nu b \bar b$ searches, however, significant event yield from $WH$ channel remains 
due to leptonic decay of the $W$ where the lepton escapes identification. 
Additionally, angular cuts between the jets and the~\MET~ are used 
to reject further the background, and identify a QCD multijets control sample,
and $b$-tagging requirements are applied to the jets.
D\O~uses a boosted decision tree (DT) technique~\cite{ref:8} trained on 25 input variables as the final 
discriminant (Figure~\ref{fig:2}). 
CDF uses two separates NN. The first one uses track-based quantities to discriminate $ZH$ from QCD multijets background.
The second one, used to extract the limit, takes as input the first NN and combines other kinematic variables to
discriminate $ZH$ and $WH$ from QCD multijet and $\ttbar$ backgrounds (Figure~\ref{fig:2}). 
In the absence of any signal, D\O~ sets a limit of 7.5 (8.4 expected) at 95\% C.L. above SM expectation, while CDF
obtains a limit of 8.0 (8.3 expected) at 95\% C.L. Both experiments have performed searches splitting the data 

\subsection{Search for $WH \ra l \nu b \bar b$}

\begin{figure}[htb]
\begin{center}
\includegraphics[height=.28\textheight]{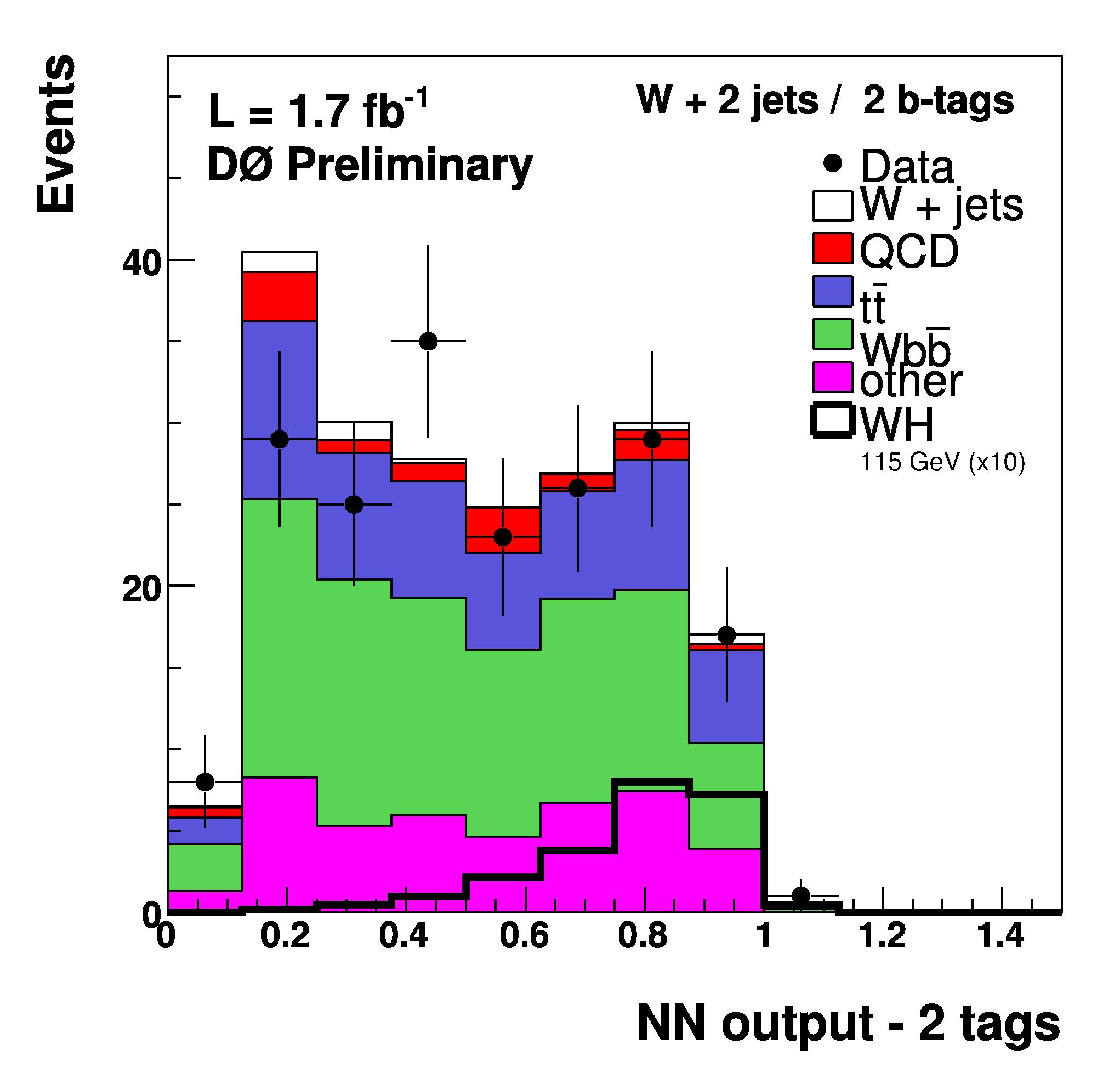}  
\includegraphics[height=.28\textheight]{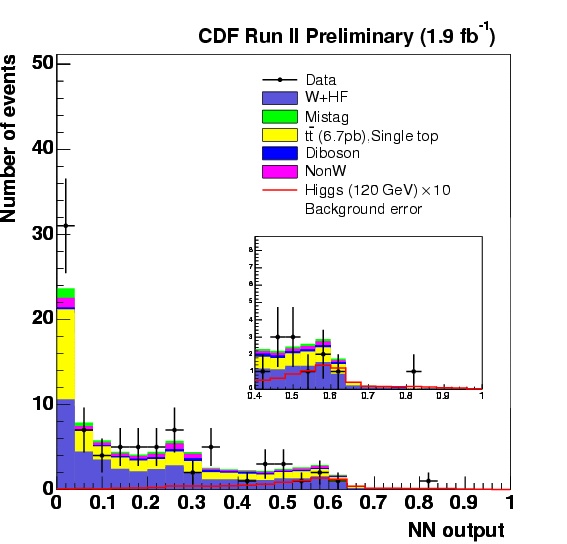}  
\caption{D\O~NN output (left) and CDF NN output (right) for the double tag sample.}
\label{fig:3}
\end{center}
\end{figure}

Searches for Higgs production in associated production with a $W$ boson decaying leptonically provides 
the most stringent constraints on the low mass SM Higgs. The event selection consists on identifying one isolated
high-$P_T$ electron or muon, \MET$>20~\GeV$ and two (two or three for D\O) jets with $E_T>20~\GeV$. 
Additionally, CDF used the ~\MET+jets trigger to identify events with one isolated track which failed the
electron or muon identification, thus increasing the Higgs acceptance by 25\%.
D\O~uses a NN trained on seven kinematic variables and splits the data into eight exclusive sets based on the 
number of $b$-tagged jets, lepton flavor and the two data periods\footnote{D\O~detector was upgraded during the 
Spring 06 Tevatron shutdown. Run IIa dataset corresponds to $~1.04~\rm{fb}^{-1}$, while
Run IIb corresponds to $~0.63~\rm{fb}^{-1}$}
to optimize the sensitivity (Figure~\ref{fig:3}). 
As no excess is observed compared to the expectation, a limit is derived from the 
eight individual analysis and combined. D\O~sets a limit of 10.9 (8.9 expected) at 95\% C.L. above SM expectation.
CDF uses a NN to improve the purity of the single tag sample based on the secondary vertex tagger.
In addition, CDF uses a jet probability tagging algorithm which identifies $b$-jet by requiring a low probability
that the tracks contained in the jets originated from the primary vertex. An event is considered double-tagged if it
contains either two secondary vertex tags or one secondary vertex tag and a jet probability tag. Note that the 
isolated track channel does not make use of neither the NN tagger nor jet probability tagger.
CDF uses a NN trained on six kinematics variables and performs the analysis separately on the eight categories of events 
(Figure~\ref{fig:3}).
In the absence of signal, CDF sets a limit of 6.4 (6.4 expected)  95\% C.L. above SM expectation.

\subsection{Search for $VH,  VBF, H \ra \tau^+ \tau^- + 2 \rm{~jets} $}

CDF performs a novel simultaneous search using the $\tau$ decay mode of the SM Higgs boson in
$W(\ra q\bar q')H(\ra \tau^+\tau^-)$, $Z(\ra \qqbar)H(\ra \tau^+\tau^-)$, 
Vector Boson Fusion (VBF) $H\ra \tau^+\tau^-$ and $gg\ra H(\ra \tau^+\tau^-)$.
Candidates events are selected by identifying an isolated electron or muon from the $\tau$ leptonic decay, one 
hadronic $\tau$ and at least 2 jets in an event. In order to further improve the search sensitivity, 
three NN are used to discriminate the SM Higgs signals from $Z\ra \tau^+\tau^-$, top and QCD multijet backgrounds.
In the absence of signal, the minimum of the three NN is used to extract a 95\% C.L. ranging from 30 to 150 
(24 to 112 expected) for Higgs masses of $110-150~\GeV$. 
  

\subsection{Search for $WH \ra WWW^*$}

The associated production of SM Higgs with a W boson, where $H\ra W^+W^-$, is important in the intermediate 
mass region ($125-145~\GeV$), although this channel suffers from a low branching ratio. Additionally, in some models
with anomalous coupling (``fermiophobic Higgs''), $BR(H\ra WW^*)$ may be close to 100\%. The search consists in 
identifying like-sign high $P_T$ isolated leptons; one coming from $H\ra WW$, the other from the prompt W. 
CDF scans the two-dimensional plane of the $2^{\rm{nd}}$ lepton $P_T$ versus the dilepton system $P_T$, 
while D\O~performs a counting experiment. Neither experiment sees an excess in the number of events over the SM background
expectation and set limits ranging from 20 to 24 times above SM for $M_H=160~\GeV$ and $M_H=140~\GeV$ respectively.   

\subsection{Search for $H \ra W^+ W^-$}

\begin{figure}[htb]
\begin{center}
\includegraphics[width=.35\textheight]{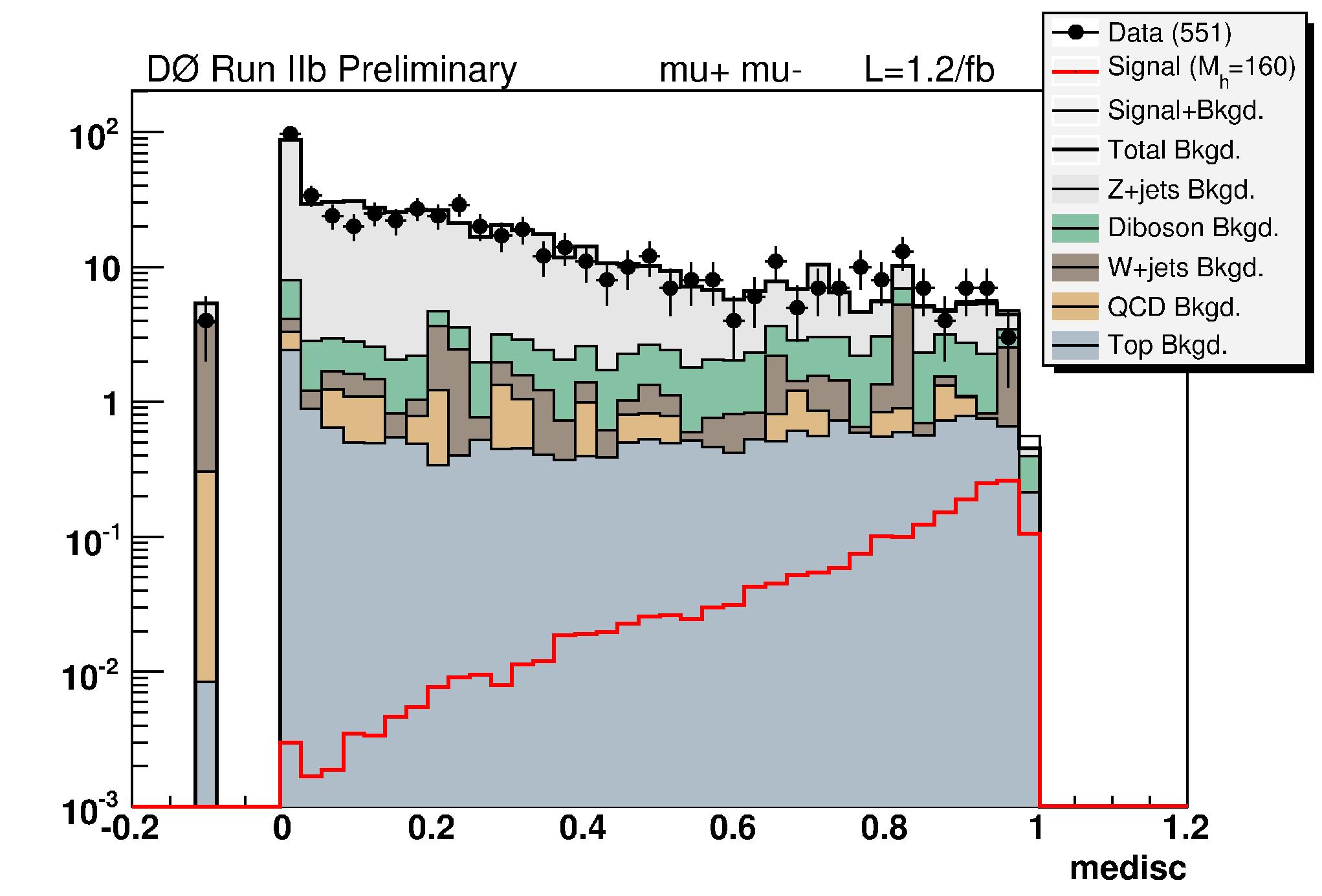}  
\includegraphics[width=.35\textheight]{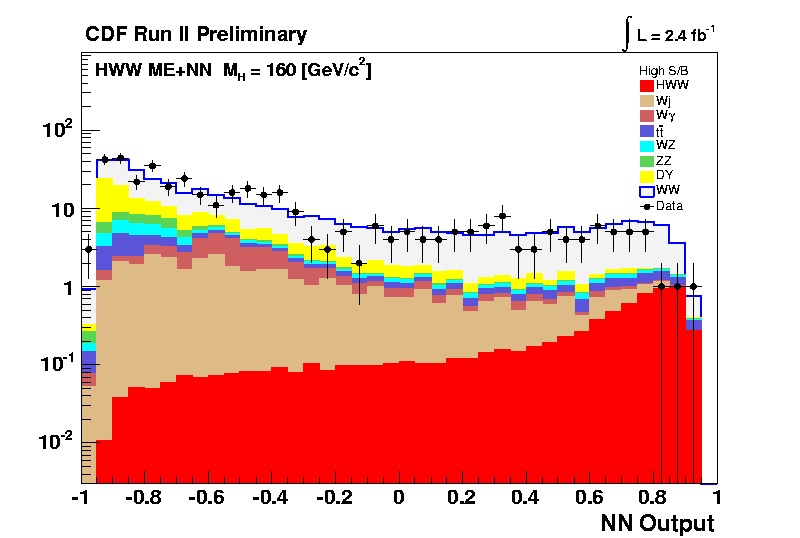}  
\caption{D\O~distribution of the ME discriminant (left) after the final selection in $\mu\mu$ channel for $M_H=160~\GeV$.
Values of $\rm{ME}_{disc}<0$ correspond to events with very small probability of either signal or WW background.
CDF NN output (right) for $M_H=160~\GeV$ for events with high S/B ratio as determined from the LR.}
\label{fig:4}
\end{center}
\end{figure}

In the high mass region (above $135~\GeV$), SM Higgs decays predominantly to $W^+W^-$. This channel benefits
from a very clean signature with low SM backgrounds thus providing the largest sensitivity for a SM Higgs boson search
at the Tevatron. The event selection consists in identifying two opposite charged isolated high $P_T$ leptons 
($ee$, $\mu\mu$ and $e\mu$), ~\MET$>20, 25$ (D\O, CDF) and little jet activities to reduce background from top pair production. 
The QCD multijet background is further reduced by requiring that the di-lepton mass be above ~$15~\GeV$. 
The remaining background is SM $WW$ production, and the opening angle between the two leptons, $\Delta\phi_{ll}$ can be used
as a discriminating variable since the leptons from a spin-0 Higgs tend to be more co-linear. D\O~tunes 
the various pre-selection criteria thresholds for each di-lepton class and the various SM Higgs masses.  
To improve the separation between signal and backgrounds, D\O~ uses a NN for each of the di-lepton channel. The input
variables consist of various event or object kinematics, angular variables and a discriminant constructed using the 
Matrix Element (ME) method in the $ee$ and $\mu\mu$ channels (Figure~\ref{fig:4}). 
D\O~obtains a 95\% C.L. of 2.1 (2.4 expected) above SM expectation for $M_H=160~\GeV$. 
In order to optimize the lepton acceptance, CDF loosen its lepton selection criteria and also identifies isolated track not
fiducial to either the calorimeter or muon chambers within $|\eta|<1.2$ thus improving the expected signal events 
by a factor of 1.6. In order to maximize the sensitivity, the ME probabilities are calculated for each selected event to be
Higgs signal or $WW$, $ZZ$, $W\gamma$, $W+\rm{jet}$ backgrounds and used to construct five likelihood ratio (LR) discriminants. 
Those LR are used in conjunction with six other kinematics variables as input to a NN (Figure~\ref{fig:4}). 
In absence of signal, the NN output is fitted to extract a 95\% C.L. of 1.6 (2.5 expected) 
above SM expectation for $M_H=160~\GeV$. 

\subsection{Combine upper limit on Standard Model Higgs boson production}

\begin{figure}[htb]
\begin{center}
\includegraphics[height=.35\textheight]{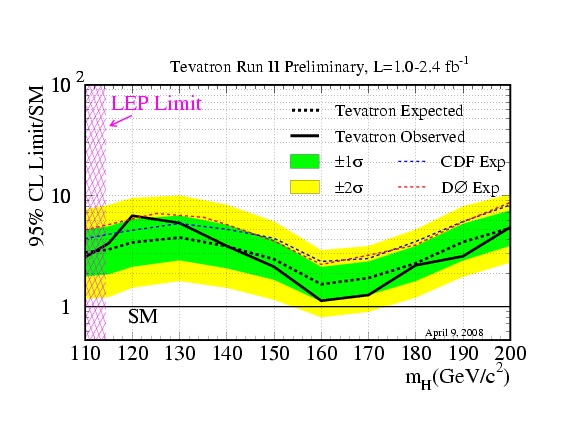}  
\caption{Observed and expected (median, for the background-only hypothesis) 95\% C.L. upper limits on the ratios to the SM 
cross section, as functions of the Higgs boson test mass, for the combined CDF and D\O~analyses. The limits are expressed 
as a multiple of the SM prediction for test masses for which both experiments have performed dedicated searches in different 
channels. The points are joined by straight lines for better readability. 
The bands indicate the 68\% and 95\% probability regions where the limits can fluctuate, in the absence of signal. Also shown 
separately are the expected upper limits obtained for all combined CDF and D\O~channels.}
\label{fig:5}
\end{center}
\end{figure}

Since no single decay channel and neither experiment has sufficient statistical power to reach the SM prediction over the 
full mass range, results from all searches for both experiments are combined. 
In order to simplify the combination, the  searches are separated into twenty nine (13 for CDF and 16 for D\O) mutually 
exclusive final states. All systematic uncertainties and their correlations between channels and across 
the experiments are taken into account to perform several types of combinations, using Bayesian and 
Modified Frequentist approaches, and found to be in agreement within 10\%. The results are presented in Table~\ref{tab:1} and 
Figure~\ref{fig:5}.

\begin{table}[tb]
\begin{center}
\begin{tabular}{l|cccccccccccc}  
\hline \hline
$m_H (\GeV)$ &  110 & 115 & 125 & 135 & 140 & 150 & 160 & 170 & 180 & 190 & 200 \\ \hline
Expected & 3.1  & 3.3 & 3.8 & 4.2 & 3.5 & 2.7 & 1.6 & 1.8 & 2.5 & 3.8 & 5.1  \\ 
Observed & 2.8  & 3.7 & 6.6 & 5.7 & 3.5 & 2.3 & 1.1 & 1.3 & 2.4 & 2.8 & 5.2 &  \\ \hline \hline
\end{tabular}
\caption{Tevatron combined 95\% C.L limits on $\sigma\times BR(H\ra \bbbar/W^+W^-)$ for 
SM Higgs boson production. The limits are reported in units of SM production cross section times branching fraction.}
\label{tab:1}
\end{center}
\end{table}

\section{Non Standard Model Higgs searches}

\subsection{Search $\phi \ra \tau^+ \tau^- $}

D\O~\cite{ref:9} and CDF have searched for a MSSM Higgs boson decaying to a tau pair, where one of the tau decays leptonically 
to an electron ($\tau_e$) or muon ($\tau_{\mu}$) and the other decay either hadronically ($\tau_h$) or leptonically but
to a different lepton flavor. Most of the sensitivity in this channel comes from the identification of $\tau_h$.
D\O's tau identification is based on three different NN aimed to optimize the background rejection depending on the hadronic 
tau types. The $\tau_h$ decays are subdivided into three types: single charged track matches the energy deposited 
in the hadronic calorimeter (consistent with $\tau^\pm\ra\pi^\pm\nu$); single charged track matches the energy deposited 
in the electromagnetic and hadronic calorimeter (consistent with $\tau^\pm\ra\pi^\pm\pi^0\nu$); three charged tracks with 
invariant mass less than $1.7~\GeV$ (consistent with $\tau^\pm\ra\pi^\pm\pi^\pm\pi^\mp(\pi^0)\nu$). 
For the second case of hadronic tau decay, if there is a significant amount of electromagnetic energy, a 
fourth NN is used to discriminate tau decays from direct electron production. 
CDF's $\tau_h$ identification starts by reconstructed a high $P_T$ ``seed'' track and then associated to the $\tau$
candidate additional tracks within a narrow $\eta-\phi$ cone. The total momentum in the isolation annulus around the 
$\tau$-cone is required to be small. The $\tau$ momentum is measured by combining the track momenta in the $\tau$-cone 
(typically originating from charged pions), with the electromagnetic calorimeter momenta (typically originating from neutral
pions), with a correction to account for the charged-pion contribution to the electromagnetic momenta.
Both D\O~and CDF exploit the visible mass, $M_{\rm{vis}}$ spectrum as a figure of merit, with 
$M_{\rm{vis}}=\sqrt{(P_{\tau_1}+P_{\tau_2}+/\!\!\!\!E_{T})^2}$ and where $P_{\tau_{1,2}}$ are the four-vector of the visible 
tau decays. 

\begin{figure}[htb]
\begin{center}
\includegraphics[width=.28\textheight]{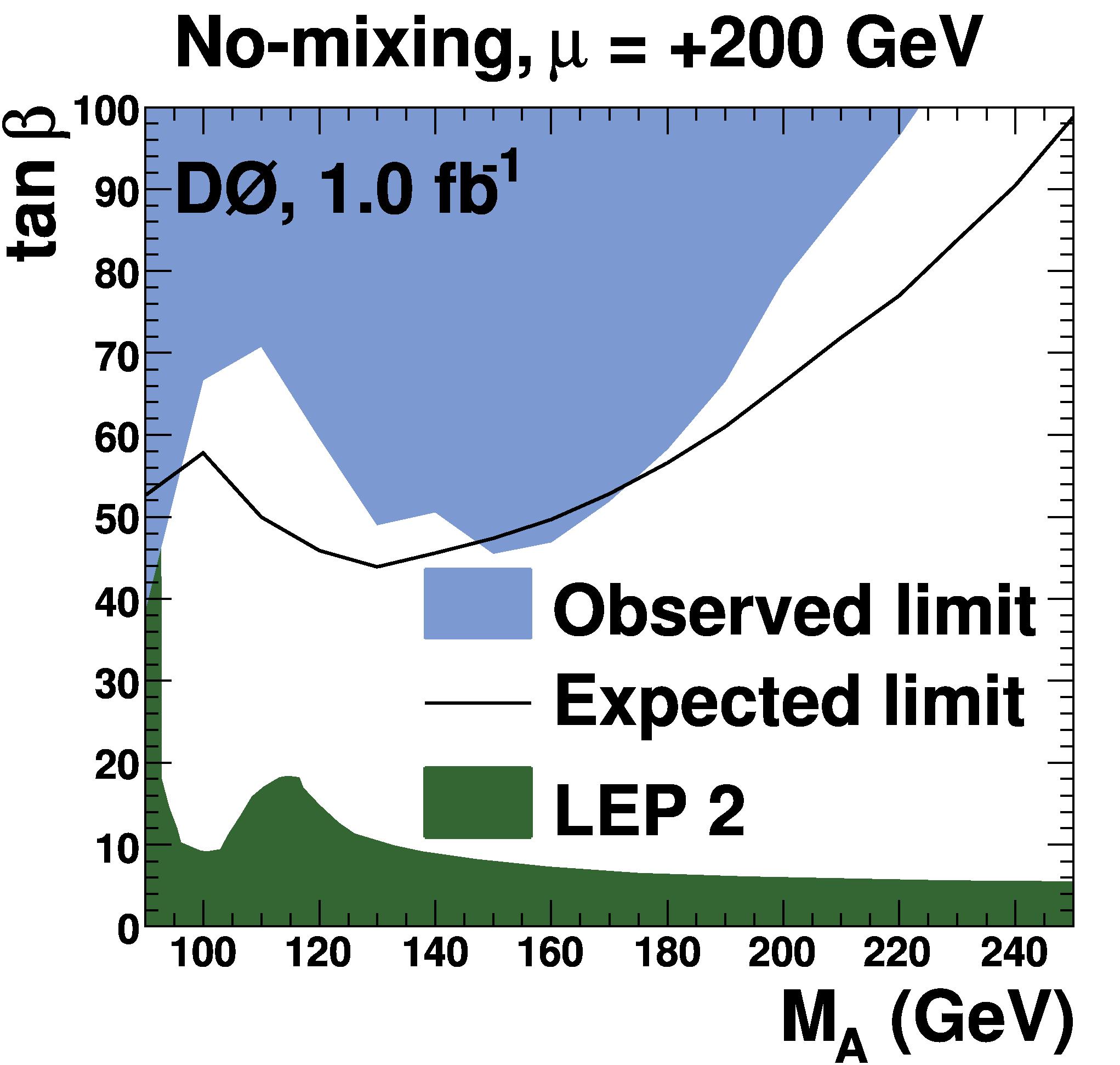}  
\includegraphics[height=.26\textheight]{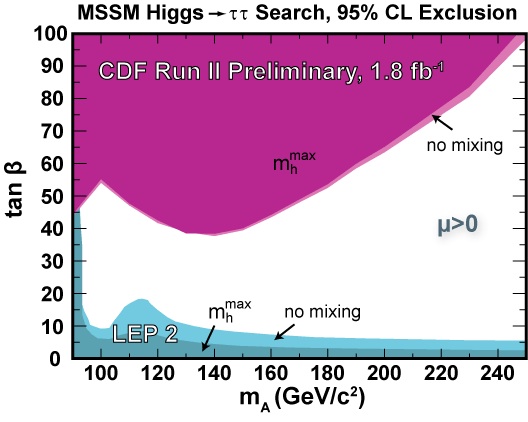}  
\caption{D\O~(left) and CDF (right) limits in the $m_A-\tan\beta$ plane. D\O~examines the no-mixing and 
$m_h^{\rm{max}}$ (not shown) scenarios for $\mu>0$. The $m_h^{\rm{max}}$ limits are similar. 
CDF examines the $m_h^{\rm{max}}$ and no-mixing for $\mu>0$ and $\mu<0$ (not shown). The $\mu<0$ limits are similar.}
\label{fig:6}
\end{center}
\end{figure}

No excess consistent with a MSSM Higgs is observed in either D\O~or CDF data, so limits in the $m_A-\tan\beta$ plane are set
for non-vanishing (``$m_h^{\rm{max}}$") or vanishing (``no-mixing'') stop mixing with $\mu>0$ and $\mu<0$, 
where $\mu$ is the Higgs mixing parameter at the electroweak scale (Figure~\ref{fig:6}).

\subsection{Search $ b \phi \ra bb \bar b $}

\begin{figure}[htb]
\begin{center}
\includegraphics[height=.45\textwidth]{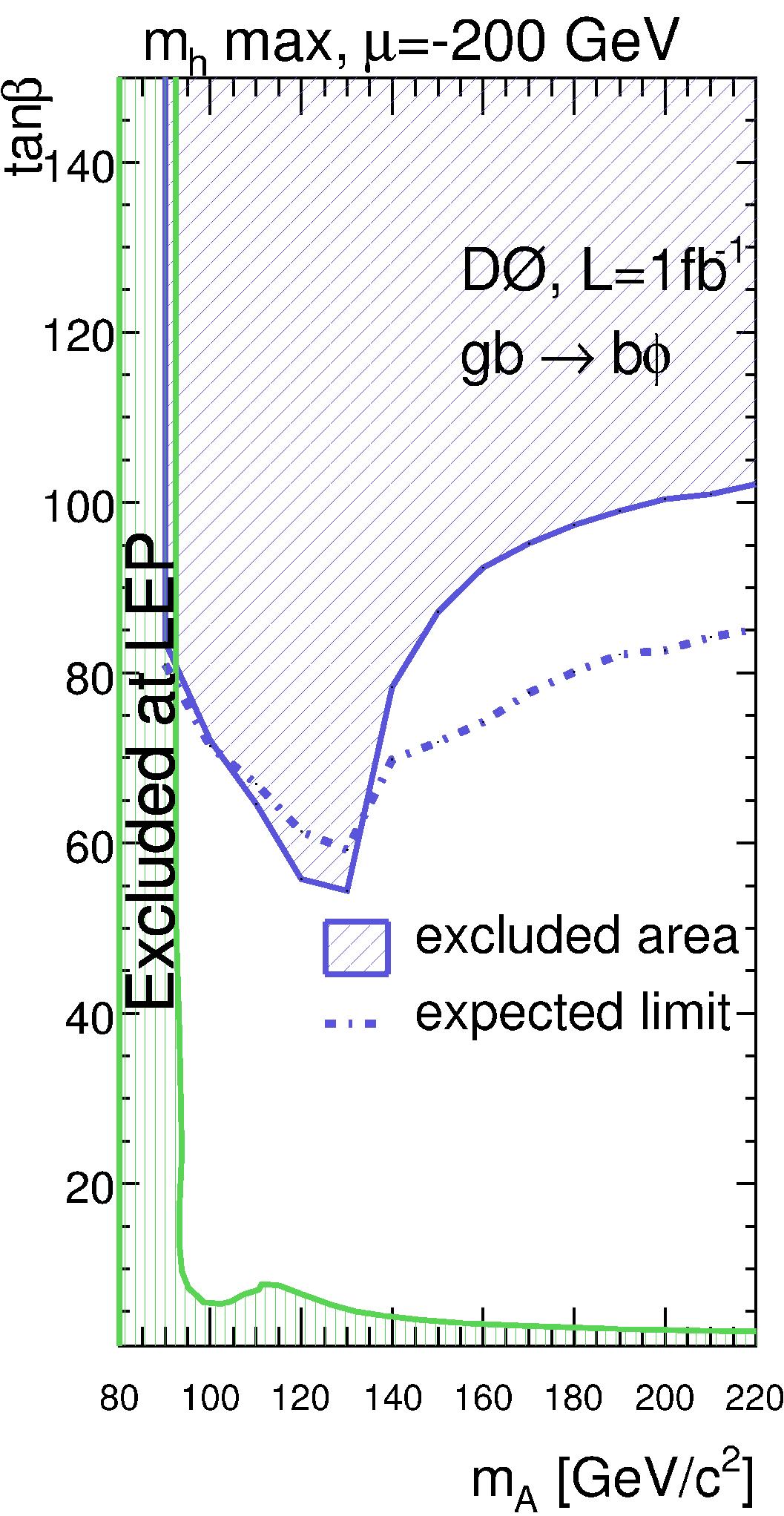}  
\includegraphics[width=.35\textheight]{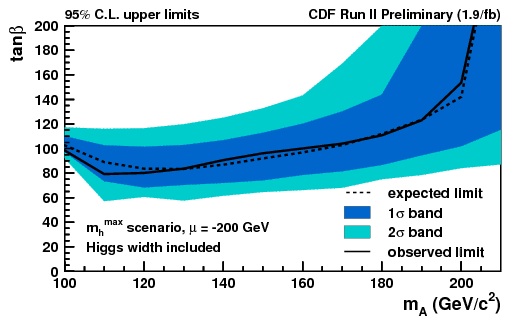}  
\caption{D\O~(left) and CDF (right) limits in the $m_A-\tan\beta$ plane. D\O~examines for $\mu\pm200~\GeV$,
no-mixing (not shown) and $m_h^{\rm{max}}$ scenarios ($\mu<0$ only).  
CDF examines the $m_h^{\rm{max}}$ scenario for $\mu=-200~\GeV$.}
\label{fig:7}
\end{center}
\end{figure}

The $\ppbar\ra H\ra\bbbar$ process is overwhelmed by direct $\bbbar$ production, therefore searches
in this Higgs decay mode use the process $gb\ra Hb\ra bbb$ and $\qqbar/gg\ra bbH\ra bbbb$. Both D\O~and CDF
require three identified $b$-jets in the final state. The $b$-jet energy resolution is important to reconstruct the
Higgs mass from the background continuum. The dominant background are QCD multijet events with two real $b$-quarks and 
a ``$b$'', ``$c$'' or ``fake'' tag. 
D\O~uses a NN $b$-tagger to identify $b$-jets. The background is estimated separately for three, four and five-jets
channels using the measured $b$-jet identification efficiency and control samples with and without identified $b$-jets.
A likelihood based on six variables is used to discriminate Higgs signal from background, where the cut on 
the likelihood varies depending on the jet multiplicity and the Higgs boson mass. The resulting invariant mass spectrum 
from the two leading jets shows good agreement between data and predicted background. Limits are set in the
$m_A-\tan\beta$ plane for $\mu\pm200$ for the $m_h^{\rm{max}}$ (negative $\mu$ only) and no-mixing scenarios 
(Figure~\ref{fig:7})~\cite{ref:10}.
CDF identifies b-jets using the secondary vertex tagger and the invariant mass of the reconstructed vertex since 
the vertex mass is higher for b-jet than for light-flavor jet. The mass distribution for four background types
($bbb$, $bbx$, $bcb$, $bqb$) are predicted using a combination of data and PYTHIA Monte Carlo simulation.
Those background templates are used in a binned maximum-likelihood to fit the data without using any absolute normalisation.  
The invariant mass spectrum of the two leading jets shows no deviations from the expected background only hypothesis 
and limits are set in the  $m_A-\tan\beta$ plane for $\mu=-200~\GeV$ in the $m_h^{\rm{max}}$ scenario (Figure~\ref{fig:7}).

\subsection{Search for Fermiophobic and doubly charged Higgs}

\begin{figure}[htb]
\begin{center}
\includegraphics[width=.32\textheight]{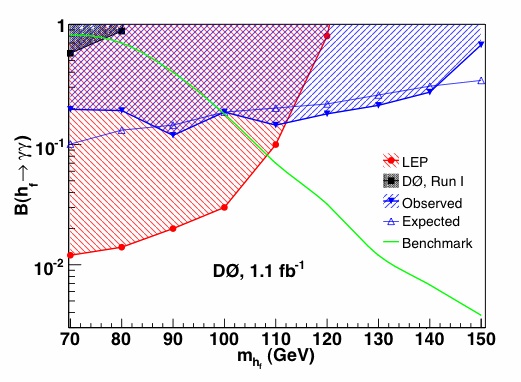}  
\includegraphics[width=.28\textheight]{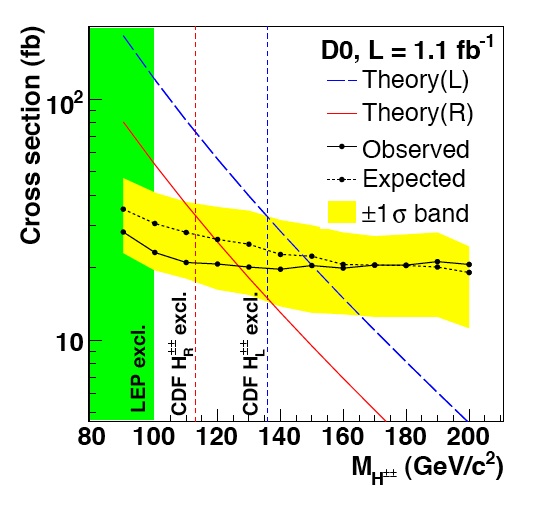}  
\caption{Limits on $BR(h_f\ra \gamma \gamma)$ as a function of the Higgs mass (left). Shaded region correspond to 
the excluded values of the branching ratio.
Cross section limit as a function of $m_{H^{\pm\pm}}$ at 95\% C.L (right).}
\label{fig:8}
\end{center}
\end{figure}

D\O~searched for fermiophobic Higgs boson produced by vector-boson fusion or in association with a $W$ or $Z$ boson, 
where for $m_H<100~\GeV$, the Higgs boson predominantly decays to photon pairs~\cite{ref:11}. 
The search consists in identifying two
high $E_T$ isolated photon. The dominant background come from direct photon production or events where jets are 
misidentified as photons, and are estimated using a combination of data and MC simulation. Since no excess of 
events is observed in the di-photon invariant mass, a 95\% C.L. upper limit on the branching ratio of a 
fermiophobic Higgs to di-photon is set (Figure~\ref{fig:8}). This search significantly extends sensitivity into region not 
accessible at LEP. 
  
Scenarios such as left-right symmetry models, Higgs triplet models and 
Little Higgs models, predict the existence of doubly-charged Higgs ($H^{\pm\pm}$) which couples to 
like-sign dilepton pairs.
At the Tevatron, doubly-charged Higgs are produced predominantly in pairs. Events with four lepton in the final state
have negligible background and therefore searches typically require fewer than four leptons to increase the acceptance.
D\O~has performed a search in the three muon final state~\cite{ref:12}, where at least two muons are like-sign. 
Since no excess is 
observed in the same sign dimuon invariant mass, the distribution is used to extract a 95\% C.L upper limit on the 
doubly charged Higgs production cross section (Figure~\ref{fig:8}). 
The limit excludes $m_{H^{\pm\pm}}>150~(127)~\GeV$ for Higgs bosons couple to left-(right-) handed muons
with 100\% branching ratio. This result increases the previous mass limit obtained at LEP and complements CDF.

\section{Conclusions and prospects}

CDF and D\O~ have performed searches for the SM and non-SM Higgs bosons over a wide range of masses with an integrated
luminosity up to $2.4\rm{fb}^{-1}$. 
Since no excess of signal above the expected backgrounds were observed, limits were set.

Both experiments have brought a variety of improvements to the analyses: 
trigger and lepton identification optimisation, dijet mass resolution, $b$-tagging algorithms, splitting classes of events, 
advance analysis techniques (NN, DT, ME), all of which permitted to improve on the limit faster than the gain from increasing
luminosity. 
In the case of a SM Higgs, a Tevatron combined limit was produced and shows that exclusion of a SM Higgs around $160~\GeV$
is around the corner.   

The search for non-SM Higgs bosons show very promising sensitivity and have already produce powerful limits. 
Analysis techniques are well advanced but additional improvements could help obtained even more stringent limits.
Additionally, both experiments plan on working toward combining their results from the different channels. 
With the expected $6-7\rm{fb}^{-1}$ of data per experiment by the end of Run II, it is expected the Tevatron
could probe down to $\tan\beta$ around 20 for low $m_A$.

\paragraph
\noindent
I would like to thank my colleagues from the D\O~ and CDF Collaborations for providing the material for this 
talk and the organisers of PIC 2008 for a very enjoyable and interesting conference. 

\bigskip

\end{document}

%% file: econfmacros.tex
\def\beq{\begin{equation}}
\def\eeq#1{\label{#1}\end{equation}}
\def\eeqn{\end{equation}}

\def\beqa{\begin{eqnarray}}
\def\eeqa#1{\label{#1}\end{eqnarray}}
\def\eeqan{\end{eqnarray}}

\let\bar=\overbar

\def\Dslash{\not{\hbox{\kern-4pt $D$}}}
\def\dslash{\not{\hbox{\kern-2pt $\del$}}}

\def\msb{{\bar{\ssstyle M \kern -1pt S}}}

